\documentclass[useAMS,usenatbib]{mn2e}
\usepackage{graphicx}

\usepackage[utf8]{inputenc}

\title[Optical counterparts of two ULXs]
{Optical counterparts of two ultraluminous X-ray sources NGC\,4559 X-10 and NGC\,4395 ULX-1}
\author[Vinokurov, Fabrika \& Atapin]
  {A.~Vinokurov,$^{1}$\thanks{vinokurov@sao.ru}
  S.~Fabrika$^{1,2}$
  K.~Atapin$^{3,1}$\\
  $^1$ Special Astrophysical Observatory, Nizhnij Arkhyz 369167, Russia\\
  $^2$ Kazan Federal University, Kremlevskaya 18, Kazan 420008, Russia\\
  $^3$ Sternberg Astronomical Institute, Moscow State University, Universitetsky pr., 13, Moscow 119991, Russia
  }

\pagerange{\pageref{firstpage}--\pageref{lastpage}}
\pubyear{2016}

\def\LaTeX{L\kern-.36em\raise.3ex\hbox{a}\kern-.15em
    T\kern-.1667em\lower.7ex\hbox{E}\kern-.125emX}

\begin{document}

\label{firstpage}

\maketitle

\begin{abstract}

We study the optical counterparts of ultraluminous X-ray sources NGC\,4559 X-10 and NGC\,4395 ULX-1. Their absolute magnitudes, after taking the reddening into account, are $M_V \approx -5.3$ and $M_V \approx -6.2$, respectively. The spectral energy distribution of the NGC\,4559 X-10 counterpart is well fitted by a spectrum of an F-type star, whereas NGC\,4395 ULX-1 has a blue power-law spectrum. Optical spectroscopy of NGC\,4395 ULX-1 has shown a broad and variable HeII~$\lambda$4686 emission, which puts this object in line with all the other spectrally-studied ULXs. Using the {\it Swift} archival X-ray data for NGC\,4395 ULX-1, we have found a period of $62.8\pm 2.3$\, days. The X-ray phase curve of the source is very similar to the precession curve of SS\,433. The optical variation of the counterpart (between two accurate measurements) amounts to 0.10 mag. Analyzing the absolute magnitudes of 16 well-studied ULX counterparts one may suggest that as the original accretion rate decreases (but nevertheless remains supercritical), the optical luminosity of the wind becomes dimmer and the donor star dominates. However, an observational bias may also influence the distribution.

\end{abstract}


\section{Introduction}

Ultraluminous X-ray sources (ULXs) are variable, non-nuclear X-ray sources in external galaxies with isotropic luminosities higher than $\sim 2 \times 10^{39}$ erg/s, which exceeds the Eddington limit for a typical stellar-mass black hole \citep{Feng2011}. There are two popular models to explain the ULX phenomenon. One involves intermediate-mass black holes (IMBHs) of $10^2 - 10^4 M_{\odot}$ with standard accretion discs \citep{Colbert1999}; the second model considers stellar-mass black holes ($\sim 10 M_{\odot}$) accreting at super-Eddington rates \citep{Poutanen2007}.

X-ray studies of ULXs have shown that the behavior and shape of the spectra of these objects differ strongly from what is observed in galactic black holes. The often observed high-energy curvatures \citep{Stobbart2006,Gladstone2009,Caballero2010} with a downturn between $\sim 4$ and $\sim 7$ keV in the X-ray spectra of ULXs suggest that the ULX accretion discs are not standard. The inner parts of the accretion discs may be obscured by a hot outflow or optically thick corona \citep{Gladstone2009} which comptonize radiation from the inner disc.

Observations of ULXs and their environment in the optical range provide additional information about the objects themselves, e.g. about the masses of their progenitor stars which turn out to be greater than 50 solar masses \citep{Poutanen2013}. Observations of nebulae surrounding ULXs \citep{Pakull2002,Lehmann2005,Abolmasov2007a,Kaaret2010} testify that they form due to jets or powerful winds. All ULXs identified in the optical range (about 20 objects are identified reliably) are faint sources with m$_V = 21 - 24$ \citep{Tao2011}, the brightest counterpart being ULX P13 in NGC\,7793, m$_V \approx 20.5$ \citep{Motch2014}.
 
At present, the spectra of less than 10 optical counterparts of ULXs have been studied. \cite{Fabrika2015} have shown that optical spectra of ULXs contain broad emission lines of HeII~$\lambda4686$ and of hydrogen H$_{\alpha}$ and H$_{\beta}$
 with $FWHM \sim 1000$ km/s. All the optical spectra turned out to be similar to each other (see also \cite{Roberts2011,Cseh2011,Motch2014}), and to the 
spectra of WNLh-type  stars \citep{Sholukhova2011} and SS\,433 \citep{Fabrika1997,Fabrika2004}. In the mentioned paper it was also assumed that the studied ULXs represent a uniform class of objects that are probably supercritical accretion discs. Recent spectroscopy of M81 ULS-1 \citep{Liu2015} has shown, in addition to the broad HeII, H$\beta$, and H$\alpha$ lines, the presence of blueshifted and redshifted H$\alpha$ lines forming in baryonic relativistic jets. Previously, the relativistic lines were observed only in SS\,433 \citep{Fabrika2004}.

Here we present the identification of the optical counterpart of NGC\,4559 X-10 and accurate astrometry for NGC\,4395 ULX-1. The first source with the X-ray luminosity $L_X \sim 7 \times 10^{39}$ erg s-1 is located in a star-forming region of a late-type spiral galaxy at a distance of 7.3 Mpc \citep{Tully2013}. X-10 was studied in the optical range by \cite{Cropper2004} and \cite{Ptak2006}, but they were unable to find an unambiguous identification for this object. The luminosity of NGC\,4395 ULX-1 is about $4 \times 10^{39}$ erg/s in its bright state. The source is located in a nearby Seyfert galaxy at a distance of 4.76 Mpc \citep{Tully2013}. The optical counterpart of NGC\,4395 ULX-1 was first identified in \cite{Gladstone2013}. We also report the presence of a long-term X-ray variability in NGC\,4395 ULX-1 and present its optical spectroscopy. We analyze the spectral energy distributions of both ULXs using the {\it Hubble Space Telescope} ({\it HST}) data and discuss the optical luminosities of these two sources in comparison with other well-known ULXs.

\section{Observations and Analysis}

\subsection{Astrometry and optical counterparts}

The archive images from {\it Chandra X-Ray Observatory} and {\it HST} were used to identify the optical counterparts of NGC\,4559 X-10 and NGC\,4395 ULX-1; reference sources were used to improve the relative astrometry.   

In the case of NGC\,4559 X-10, the reference object was the well known ULX NGC\,4559 X-7 (e.g., \cite{Soria2005,Tao2011}). Both sources are located on chip S3 of ACIS with a moderate offset from the optical axis (less than 1.9$\arcmin$) in the {\it Chandra} observation (ID 2026). The best angular resolution {\it HST} observation of the X-10 region was taken on March 9, 2005, with ACS/HRC in the F555W filter. Since X-10 and X-7 are located in different images in all {\it HST} observations, for X-7 we chose an ACS/WFC/F550M image taken on the same date. 

In order to correct for the offset in coordinates between the {\it HST} observations of these two sources we used a {\it g}-band SDSS (Sloan Digital Sky Survey; \cite{Alam2015SDSS}) image. For the offset between the {\it HST} images of X-10 and SDSS we used two additional reference stars and a bright point-like optical source in the nuclear region of the galaxy. To determine the shift between the {\it HST} image of X-7 and the SDSS image we used four bright isolated stars near X-7. Finally, using the corrected position of the X-7 counterpart in the optical and {\it Chandra} images we derived a position for X-10 on ACS/HRC/F555W R.A. = $12^{h}$ $35^{m}$ $58^{s}.512$, Dec = $+$ $27^{\circ}$ $57\arcmin$ $42\arcsec.87$ (J2000.0) with a 1$\sigma$-accuracy of 0.15$\arcsec$. 

To make astrometric measurements for NGC\,4395 ULX-1, we chose the {\it HST} observation taken on March 31, 2014, with WFC3/UVIS in the F438W filter. There is only one {\it Chandra} observation of NGC\,4395 ULX-1 (ID 402) with a large offset from the optical axis of 4.5$\arcmin$. Due to a considerable shift along the optical axis, the PSF shape of the object is strongly distorted, which leads to the relatively low accuracy of its coordinate measurements. Three X-ray sources from {\it Chandra} which we identified in the SDSS image were used as reference sources. The corrected position of ULX-1 relative to {\it HST} is R.A. = $12^{h}$ $26^{m}$ $01^{s}.437$, Dec = $+$ $33^{\circ}$ $31\arcmin$ $31\arcsec.18$ with an accuracy of about 0.3$\arcsec$. 

In Fig.\,1 we present the positions of our ULXs. There is a single relatively bright object in the {\it HST} image within the corrected {\it Chandra} X-ray error box of NGC\,4559 X-10; a much fainter feature lies near the edge of the error circle. The bright source has diffuse morphology with a sharp brightening toward the center, its size being  $\simeq 0.09\arcsec \times 0.08\arcsec$, whereas the surrounding stars have a full width at half maximum of $FWHM \simeq 0.05\arcsec$. Apparently, this source is a stellar-like object surrounded by faint unresolvable stars. The only optical counterpart of NGC\,4395 ULX-1 within the error box of the X-ray coordinates is a stellar-like object. 

\begin{figure*}
\label{Fig1}
\begin{center}
\includegraphics[angle=0,scale=0.40]{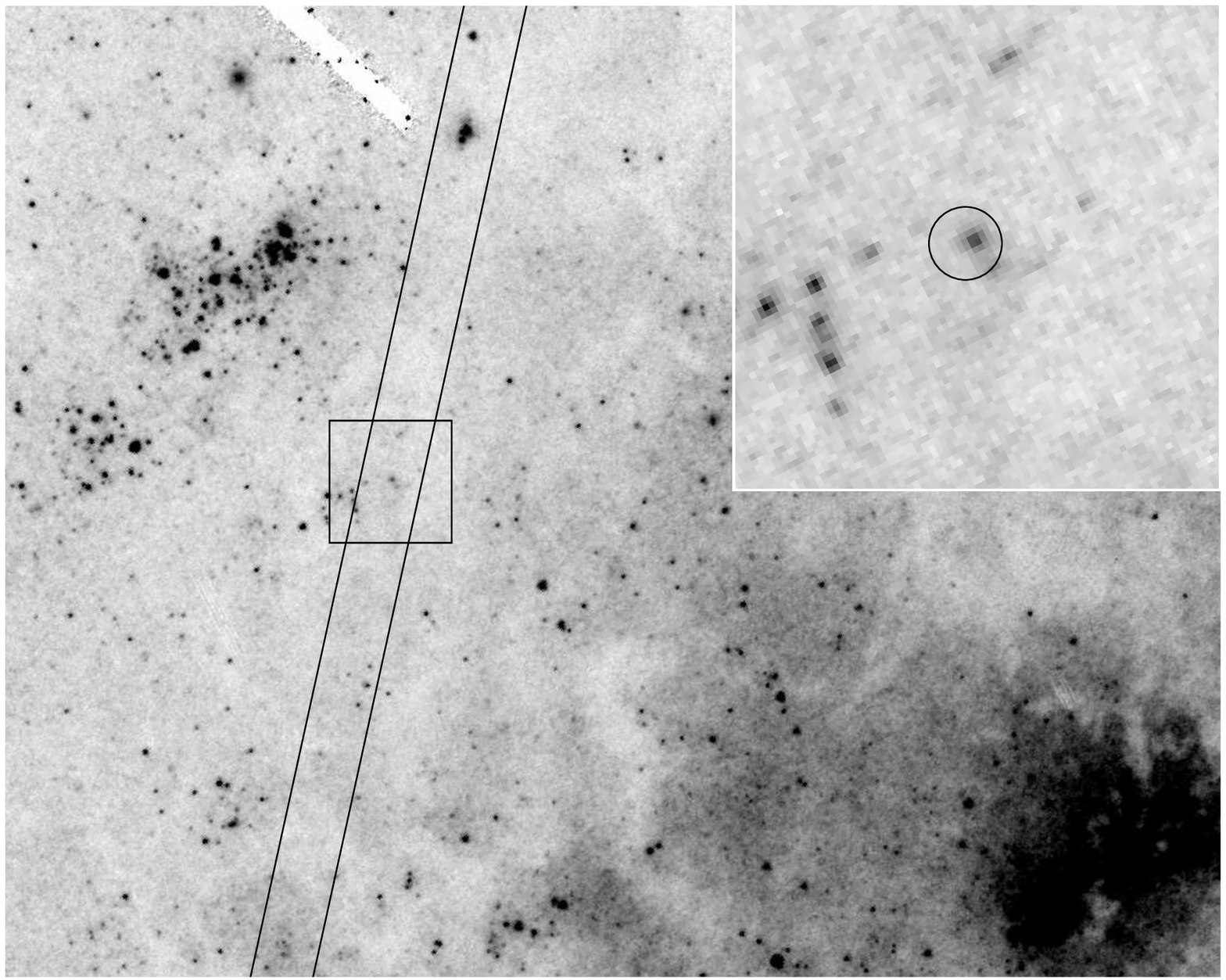}
\includegraphics[angle=0,scale=0.40]{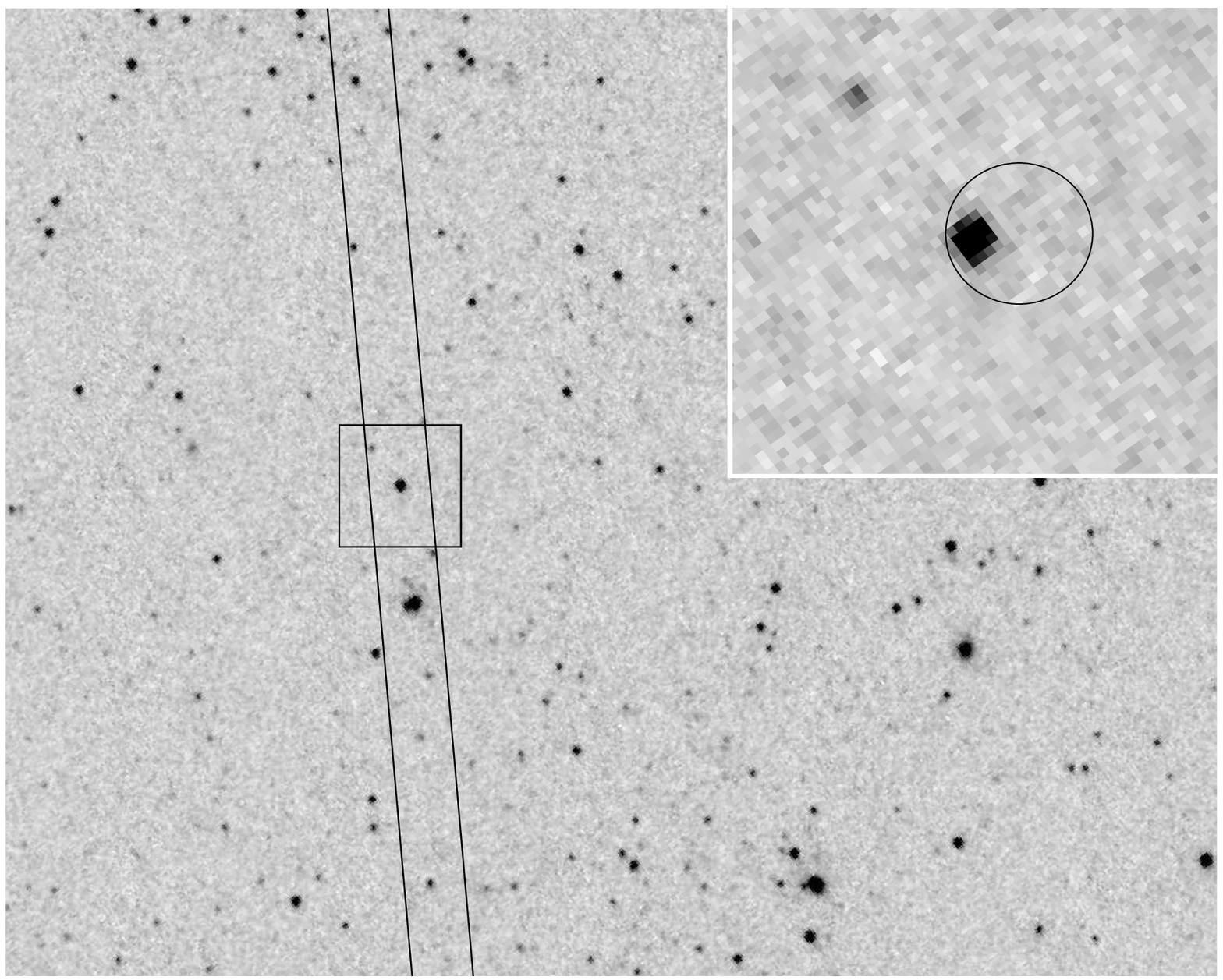}
\caption{{\it HST} images in F555W and F438W around NGC\,4559 X-10 (left) and NGC\,4395 ULX-1 (right), respectively. The zoomed images show square regions with a size of 2$\arcsec$, where the 0.15$\arcsec$ (left) and 0.3$\arcsec$ (right) radii circles indicate the astrometrically corrected X-ray boxes. North is top, East is left. On both images we show the 1$\arcsec$ slits capturing the X-10 region and the optical counterpart of ULX-1.
}
\end{center}
\end{figure*}

\subsection{Photometry and spectral energy distributions of the optical counterparts}

To study the spectral energy distributions (SEDs) in the optical range we used ACS/HRC images in F435W, F555W, F814W for NGC\,4559 X-10, and WFC3/UVIS/F275W, F336W, and F438W images for NGC\,4395 ULX-1 from the same {\it HST} datasets as for astrometry. Photometry was performed on drizzled images, using the {\scshape apphot} package in {\scshape iraf}. All magnitudes are given in the Vegamag system. The background was estimated from a concentric annulus around the objects.

We have performed photometry of the bright optical counterpart of NGC\,4559 X-10 and the faint source near the error circle boundary (Fig.\,1). To reduce the contribution of faint stars surrounding the X-10 counterpart, we chose a small aperture with a 2-pixel radius (0.05$\arcsec$). Aperture corrections were calculated using the {\scshape CALCPHOT} procedure of the SYNPHOT package. We were unable to measure the aperture corrections directly because of the small field of view of ACS/HRC and the high density of stars in this region. The reddening correction was carried out in SYNPHOT package using the extinction measured from our spectra (see below). The dereddened magnitudes are m$_{F435W} = 24.38\,\pm\,0.07$, m$_{F555W}=24.04\,\pm\,0.04$, and m$_{F814W}=23.68\,\pm\,0.04$. The fainter source at the X-10 error box boundary is identified as a relatively isolated source only in the images in filters F435W and F555W. Its dereddened magnitudes are m$_{F435W} = 25.05\,\pm\,0.12$ and m$_{F555W}=24.91\,\pm\,0.09$.

Photometry of the NGC\,4395 ULX-1 counterpart was performed in an aperture with a 3 pixel radius. Aperture corrections for stellar magnitudes in each filter were determined by 3$-$5 bright stars. Extinction in the optical range was determined from spectroscopy of a nebula around the object. The extinction-corrected stellar magnitudes of the ULXs are equal to m$_{F275W} = 19.971\,\pm\,0.015$, m$_{F336W}=20.497\,\pm\,0.016$, and m$_{F438W}=22.075\,\pm\,0.016$. The stellar magnitude errors for all objects do not include the uncertainty related to reddening.   

The filter wavelengths were corrected for the spectral slope; they were calculated using the CALCPHOT task with the {\it avglam} parameter in the SYNPHOT package. The fluxes of the ULX-1 counterpart are in good agreement with the power law $F_\lambda \propto \lambda^{-\alpha}$, where $\alpha = 2.87 \pm 0.07$. On the contrary, the optical SED of X-10 is not power-law but may be fitted by a spectrum of an F6-F8 supergiant.

To test the optical variability of NGC\,4395 ULX-1, we used additional photometry taken on March 22, 2015, with WFC3/UVIS/F547M. The dereddened magnitude is m$_{F547M} = 22.26\,\pm\,0.03$. To compare this value with the one previously obtained in March 2014 in F438W, both magnitudes were converted to Johnson V-band using SYNPHOT.  We found V$ = 22.25\,\pm\,0.03$ for March 2015 and V$ = 22.15\,\pm\,0.02$ for March 2014. Thus, we reliably detect a 0.1 magnitude variability of the object on a time scale of about a year.

During our spectroscopic observations of ULX-1 with the Russian 6 m BTA telescope we also obtained V-band images. The source magnitudes for our three best-seeing observations are $22.14\,\pm\,0.10$ (January 2014), $22.18\,\pm\,0.12$ (January 2015), and $22.15\,\pm\,0.12$ (February 2015). All magnitudes are close to those of {\it HST}.

\subsection{Spectroscopy}

Long-slit spectroscopy of both ULXs was obtained using the BTA telescope with the SCORPIO spectrograph \citep{Afanasiev2005}. All data reduction and calibrations were performed with {\scshape midas} procedures.  

For NGC\,4559 X-10, we have data obtained on December 17, 2015, with a 13~\AA \,resolution in the 3500--7200~\AA\AA\ spectral range. The seeing was 1.5\arcsec. The optical counterpart of NGC\,4559 X-10 is an extremely faint source and we managed to obtain only a spectrum of its environment. From the spectra of its nearest nebulae we determined the reddening value. Using the ratios of the nebula lines H$\alpha$, H$\beta$, H$\gamma$, and H$\delta$ we found consistent estimates, $E(B-V) = 0.26 \pm 0.06$.

NGC\,4395 ULX-1 was observed on January 1, 2014 and January 17, 2015 in a spectral range of 4000--5700~\AA\AA \, and on February 21, 2015 in a spectral range of 3600--5400~\AA\AA. The resolution was 5~\AA, the seeing was $\approx 1\arcsec$. We also observed ULX-1 and its environment on March 11, 2016 with the same mode as for X-10. Measuring the extinction value for NGC\,4395 ULX-1 is complicated due to the low brightness of the nebula lines surrounding the object and a patchy background. For NGC\,4395 ULX-1 we obtained the reddening $E(B-V) = 0.23 \pm 0.13$.

The normalized optical spectra of NGC\,4395 ULX-1 obtained in 2014 and 2015 are shown in Fig.\,2. The narrow hydrogen spectral lines H$\beta$, H$\gamma$, and the [OIII]~$\lambda\lambda$4959,5007 lines belong to a compact nebula surrounding the source. The profile of the HeII~$\lambda$4686 line appears to be variable. In the spectra obtained in January 2014 and 2015 we found a broad component of HeII with average width $FWHM \approx 700$ km/s. In February 2015 the broad line component was not detected: the line is well fitted by a Gaussian profile with $FWHM = 310 \pm 40$ km/s, which agrees well (within the errors) with the spectral resolution.

\begin{figure}
\label{Fig2}
\begin{center}
\includegraphics[scale=0.40]{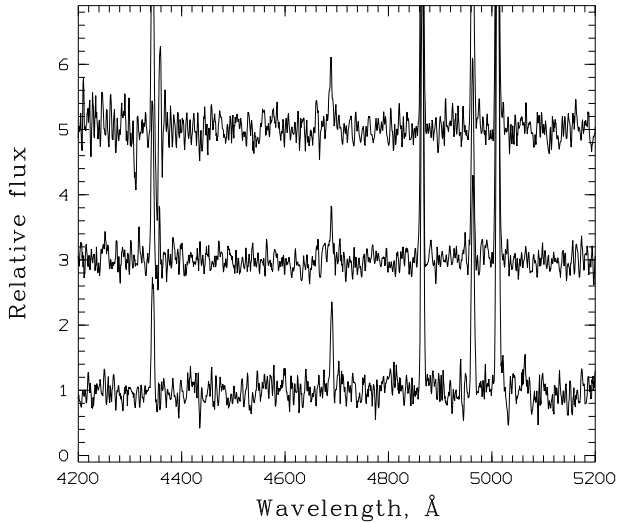}
\caption{Normalized spectra of the NGC\,4395 ULX-1 optical counterpart for the following observation dates (from top to bottom): January 1, 2014, January 17, 2015, and February 21, 2015. The strongest lines are: HeII~$\lambda$4686, the hydrogen lines H$_{\gamma}~\lambda$4340 and H$_{\beta}~\lambda$4861, and the oxygen lines [OIII]~$\lambda\lambda$4959,\,5007. In two upper spectra the broad component of the HeII~$\lambda$4686 line is clearly detected.
}
\end{center}
\end{figure}

\subsection{X-ray variability of NGC\,4395 ULX-1}

To test the X-ray variability of the source we used {\it Swift}/XRT observations. The {\it Swift} archive contains a total of 226 data sets obtained between December 2005 and April 2015, however, most of them were obtained in 2008 and 2011. Using all the available observations we extracted light curves and spectra from a circular region with a 25$\arcsec$ radius. The background region was taken in a nearby area free of other sources. We found that the X-ray flux of NGC\,4395~ULX-1 varies from 0.008 to 0.05 cnt/s, which corresponds to the X-ray luminosities from $5.6 \times 10^{38}$ to $3.5 \times 10^{39}$~erg/s in a 0.3--10\,keV range. The upper point in the light curve in Fig.\,3 corresponds to a luminosity of $4.0 \times 10^{39}$~erg/s. The background was about $3.2\times 10^{-4}$ cnt/s during all observations.

The X-ray spectra are well fitted by the two-component model \texttt{tbabs*(dikbb+powerlaw)}, yielding the disc temperature $T_d = 0.22\pm 0.03 $\,keV and $\Gamma = 2.9 \pm 0.5$ in the bright state of the object and $T_d = 0.19\pm 0.01$\,keV and $\Gamma = 3.5 \pm 0.3$ in the faint state. We have adopted $N_H = 0.25\times 10^{22}$ cm$^{-2}$ which corresponds to the optical value $E(B-V)\approx 0.23$. We conclude that despite the notable variations in the object's luminosity (with a factor of $\sim 6$) the spectrum is unchanged.

To search for periodicity in the X-ray light curve we computed the Lomb-Scargle periodogram \citep{Lomb1976,Scargle1982}. The most prominent peak greater than the $4\sigma$ level (the false alarm probability is $3.5\times10^{-5}$, \cite{Horne1986}) corresponds to a period of $62.8\pm 2.3$\,days. The light curve folded on this period is shown in Fig.\,3. The reference epoch is MJD\,53735. 

\begin{figure}
\label{Fig3}
\begin{center}
\includegraphics[scale=1.0]{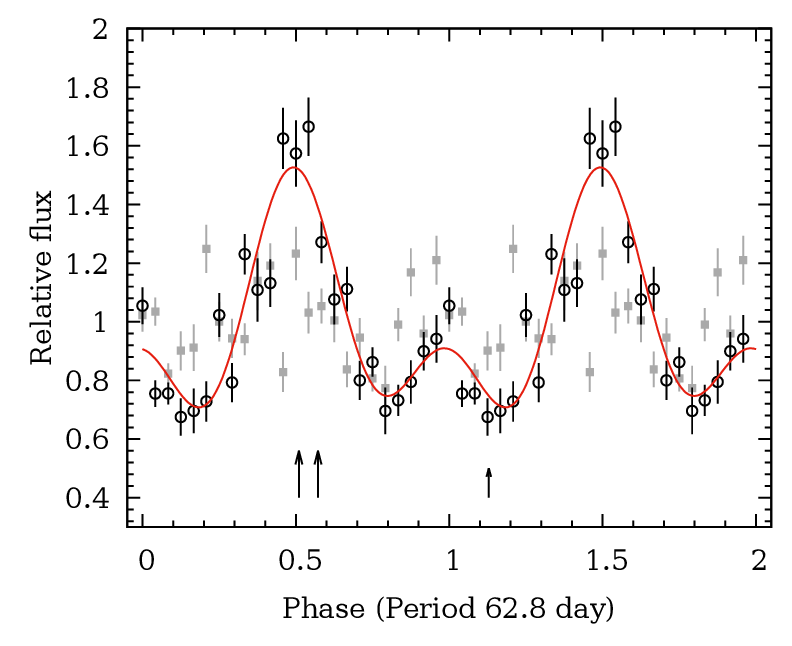}
\caption{The epoch folded X-ray light curve of NGC\,4395~ULX-1 is shown by black circles. The background folded light curve is shown by gray squares. Both light curves are in relative fluxes, the actual background flux is 50 times smaller than the object flux. The three arrows correspond to the dates when the optical spectra were taken: two longer arrows show the dates when the broad component of the HeII line was observed.
}
\end{center}
\end{figure}

\section{Discussion and Conclusions}

The spectra of all optical counterparts including NGC\,4395 ULX-1 are similar to one another. They are NGC\,1313 X-2 \citep{Roberts2011}, NGC\,5408 X-1 \citep{Cseh2011}, NGC\,7793 P-13 \citep{Motch2014}, Holmberg\,II X-1, Holmberg\,IX X-1, NGC\,4559 X-7 and NGC\,5204 X-1 \citep{Fabrika2015}, and M81 ULS-1 \citep{Liu2015}. The main feature of the spectra is the broad HeII line with FWHM $\approx 500 \div 1600$ km/s. The detection of a broad HeII line in the spectrum of the NGC\,4395 ULX-1 counterpart puts this object in line with the other ULXs and may indicate their identical nature. One may conclude that ULXs represent a roughly homogeneous class of objects, because the presence of a broad HeII emission is rare in the optical spectra. In \cite{Fabrika2015} arguments have been made as to why this type of spectrum cannot belong to a WR-type star. Recently, an ultraluminous X-ray pulsar \citep{Bachetti2014} has been detected, reaching the X-ray luminosity of $1.8 \times 10^{40}$ erg/s. Obtaining the optical spectra of this object is an important task.

The HeII line in NGC\,4395 ULX-1 has a two-component profile; the broad component was detected in only two spectra. The narrow component of the line is present in all three spectra (Fig.\,2). It might be possible that the broad component of the HeII line correlates with the X-ray light curve (Fig.\,3); in the fainter state the broad component disappears. The X-ray phase curve of NGC\,4395 ULX-1 may be connected to the precession of the supercritical disc. This is confirmed by the similarity of the phase curve and the precession curve of SS\,433 \citep{Cherepashchuk2009}. On the other hand, we did not detect a correlation between the object's optical brightness and the width of the HeII line. However, we only have three spectral observations of ULX-1, therefore the behavior of the broad component with the X-ray period phase should be confirmed by further observations.

In Fig.\,4 we find that the ULXs have a wide distribution of absolute magnitudes with a well-defined maximum at $M_V \approx -6$. The magnitudes of the optical counterparts of NGC\,4395 ULX-1 and NGC\,4559 X-10 are $-6.2$ and $-5.3$, respectively; they are also shown in the figure. The three brightest objects are SS\,433, NGC\,7793 P-13 \citep{Motch2014}, and NGC\,6946 ULX-1 \citep{Vinokurov2013,Tikhonov2014}. Other ULX counterparts in order of decreasing luminosity are: NGC\,4559 X-7, NGC\,5408 X-1, NGC\,5204 X-1, M81 ULS1, Holmberg\,II X-1, IC342 X-1, Holmberg\,IX X-1, NGC\,1313 X-2, NGC\,5474 X-1, NGC\,1313 X-1, M66 X-1 and M81 X-6 \citep{Tao2011,Tully2013,Vinokurov2013,Wu2014,Avdan2016,Tikhonov2015,Yang2011,Lee2013}. Note that the magnitudes of NGC\,4559 X-7 and NGC\,5204 X-1 were corrected for extinction ($E(B-V) = 0.10$ and $E(B-V) = 0.11$, respectively) using our previous spectroscopy of the nearby nebulae \citep{Fabrika2015}.

Out of all the objects in the diagram, three ULX counterparts, NGC\,4559 X-10, NGC\,5474 X-1, and M66 X-1 \citep{Avdan2016} have the cool spectra of F-G type supergiants, their absolute magnitudes being $M_V > -5.3$. Other objects have power-law-like spectra. The abrupt decrease in the number of objects with decreasing $M_V$ can be related both to the effects of observational selection and to the physics of the objects themselves. 

The first possibility is explained by the faintness of the objects, which makes it difficult to detect them in galaxies farther than 10 Mpc. The second possibility can be related to the decrease in the luminosity of the supercritical disc wind as the original accretion rate $\dot M$ decreases. As was shown by \cite{Fabrika2015}, the optical luminosity of supercritical discs is roughly $L_V \propto \dot M^{9/4}$, because stronger winds will reprocess more X-ray radiation emerging from the disc funnel. In the case of ULXs with the lowest optical luminosity, a considerable contribution to their luminosity can be made by donor stars. This is suggested by the cooler (on average) spectra of faint objects with $M_V > -5.3$. Accordingly, as the luminosity of the supercritical disc wind decreases, the donor star becomes dominating. 

\begin{figure}
\label{Fig4}
\begin{center}
\includegraphics[scale=0.45]{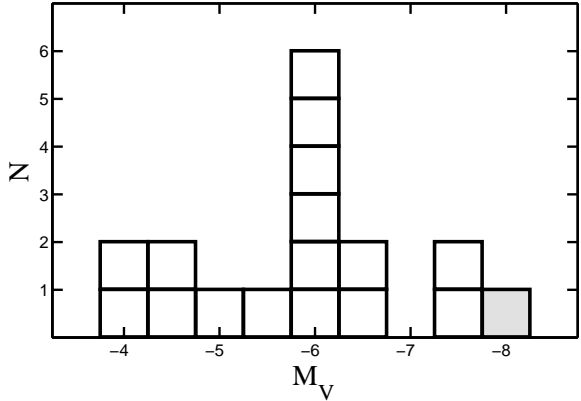}
\caption{Absolute magnitudes of well-studied ULX counterparts including our objects NGC\,4395 ULX-1 ($-6.2$) and NGC\,4559 X-10 ($-5.3$).
}
\end{center}
\end{figure}

\subsection{Acknowledgements}

\noindent
Our results are based on observations made with the NASA/ESA Hubble Space Telescope, obtained from the data archive at the Space Telescope Science Institute. STScI is operated by the Association of Universities for Research in Astronomy, Inc. under NASA contract NAS 5-26555. This research has made use of data obtained from the Chandra Data Archive and software provided by the Chandra X-ray Center (CXC) in the application package CIAO. This work has made use of data supplied by the UK Swift Science Data Centre at the University of Leicester. The research was supported by the Russian RFBR grants 16-32-00210, 16-02-00567, and the Russian Scientific Foundation grant N\,14-50-00043 for observations and data reduction. The authors are grateful to A.F. Valeev for his help with the observations.

\label{lastpage}

\end{document}